\documentclass[nofootinbib,twocolumn]{revtex4}
\usepackage{graphicx}
\usepackage{amssymb}
\usepackage{amsmath}
\usepackage{bm}
\usepackage{hyperref}
\begin{document}

\title{Non-linear dark matter collapse under diffusion}
\author{Hermano E. S. Velten}\email{velten@pq.cnpq.br}
\author{Thiago R. P. Caram\^es}\email{trpcarames@gmail.com}

\affiliation{Universidade Federal do Esp\'{\i}rito Santo (UFES), Av. Fernando Ferrari-514, 29075-910, Vit\'oria, ES - Brazil}

\begin{abstract}
Diffusion is one of the physical processes allowed for describing the large scale dark matter dynamics. At the same time, it can be seen as a possible mechanism behind the interacting cosmologies. We study the non-linear spherical ``top-hat'' collapse of dark matter which undergoes velocity diffusion into a solvent dark energy field. We show constraints on the maximum magnitude allowed for the dark matter diffusion. Our results reinforce previous analysis concerning the linear perturbation theory. \\

\textbf{Key-words}: dark matter, structure formation, velocity diffusion

PACS numbers: 98.80.-k, 95.35.+d, 95.36.+x
\end{abstract}

\maketitle

\section{Introduction}

Dark matter is a key ingredient of any viable cosmological model with a general relativity (GR) based description for gravity. Its clustering patterns are of fundamental importance to determine the observed baryonic large scale structure of galaxies and clusters. The standard description for the dark matter dynamics in the universe consists in assuming a pressureless fluid $p=0$ which fells only indirectly (via the gravitational pontential) the presence of other components like, for instance, photons, neutrinos, dark energy and baryons. This approach is based on the assumption that dark matter behaves as a perfect fluid described by the collisionless Boltzmann equation. Therefore, direct interactions with other species are in principle avoided. 

The accelerated expansion of the universe is another issue not yet properly understood. This phenomena is commonly associated to a cosmological constant $\Lambda$ which seems to be the ideal dark energy candidate. In fact, there are also other candidates for the dark energy sector which are based on either exotic fluids or scalar fields. However, no matter the actual nature of dark matter or dark energy, a primary approach in cosmology relies on the assumption that these dark components coexist without interacting directly to each other.

Due to our ignorance about the physics behind the dark components, which are the two main building blocks of modern cosmology, many possibilities about their dynamics have appeared in the literature. A notable one is the possibility that they indeed do interact in an effective way exchanging energy \cite{int}.

Many interacting cosmological scenarios have been studied in the last years. In general, the interacting term $Q$ is usually imposed into the dynamics by hand, i.e., it is said that the system dark matter plus dark energy obeys the energy balances
\begin{eqnarray}
\dot{\rho}_{dm}+3H(\rho_{dm}+p_{dm})=Q,\\
\dot{\rho}_{de}+3H(\rho_{de}+p_{de})=-Q.
\end{eqnarray}
It is worth noting that the sum of these equations correspond to a conserved dark energy sector without the interacting term.

One usually adopts phenomenological choices for $Q$ leaving the model without an explanation for the physical mechanism which drives the interaction. 

For general fluids in which the interaction of their particles with the medium is relevant the Boltzmann equation can be replaced by the Fokker-Planck equation. Mathematically, this implies in assuming a non-zero collision term in the right hand side of the Boltzmann equation. The generalization of the Fokker-Planck equation on the $4D$ curved spacetime is given by
\begin{equation}\label{FP4D}
p^{\mu}\partial_{\mu}f-\Gamma^{i}_{\mu\nu}p^{\mu\nu}\partial_{p^{i}}f=\sigma \nabla^2_p f,
\end{equation}
where $p^{\mu}$ is the four-momentum, $f$ the distribution function and $\sigma$ is the positive coefficient of matter velocity diffusion. The standard case is fully recovered when $\sigma=0$.

Matter diffusion is a physical phenomena which can take place in general relativistic universes \cite{Ca1, Ca2}. As argued in \cite{VeltenCalogero, VeltenCalogero2014} the matter diffusion present in a homogeneous and isotropic expansion - the Friedmann-Lemaitre-Robertson-Walker (FRLW) universe - can therefore be the physical mechanism behind the interaction within the dark sector. 

In Ref. \cite{VeltenCalogero} upper bounds on the magnitude of the diffusion $\sigma$ have been obtained using observational data. Background constraints from Supernovae Ia, baryonic acoustic oscillations (BAO) and the differential age of galaxies are at least one order of magnitude weaker than a first order analysis, i.e., using the cosmic microwave anisotropy spectrum and the matter power spectrum. The latter tests make use of the linear cosmological perturbation theory which probes the evolution of the density fluctuations that give rise to the large scale structure.

In this paper we go further in the study of the perturbative behaviour of the diffusion model. We investigate the non-linear stage of the perturbations, by calculating the impact of the diffusion process - here expressed by the value of $\sigma$ - on the spherical collapse. Besides, we employ the top-hat approach since no pressure gradients are present in our model. 

In the next section we describe in more detail the relativistic diffusion theory and present the equations for the FLRW background. Section III is devoted to the physics and equations of the spherical collapse where we present our results for the evolution of the decoupled collapsed region. The final section is devoted to our concluding remarks.

\section{Dark matter diffusion and its cosmological background dynamics}

As a consequence of (\ref{FP4D}) and being $T_{\mu\nu}$ and $J^\mu=n u^\mu$ respectively the energy-momentum tensor and the current density of some matter distribution in a spacetime $(M,g)$, where $n$ is the conserved particle number, matter is said to undergo {\it microscopic} (or molecular) velocity diffusion if the following kinetic equations hold

\begin{equation}\label{diffusion}
\nabla_\mu T^{\mu\nu}=\sigma J^\mu,
\end{equation}
\begin{equation}\label{consJ}
\nabla_\mu J^\mu=0.
\end{equation}
 
Equation (\ref{consJ}) assures the 4-current conservation as usual, i.e., there is no particle creation in the model. However, equation (\ref{diffusion}) modifies the standard energy-momentum balance. 

The formulation of general relativistic (GR) field equations in this new scenario has to be adapted to the condition (\ref{diffusion}). In order to keep the Bianchi identities ($\nabla_\mu T^{\mu\nu}=0$) a trivial possibility is to modify Einstein's equation as
\begin{equation}\label{EinsteinEq}
R_{\mu\nu}-\frac{1}{2}g_{\mu\nu}R+\phi g_{\mu\nu}=T_{\mu\nu},
\end{equation} 
where we have included a cosmological field $\phi$ which obeys to
\begin{equation}\label{phieqfluid}
\nabla_\mu\phi=\sigma n u_\mu.
\end{equation}
Note we have used used units $8\pi G=c=1$.  

As usual, $R_{\mu\nu}$ denotes the Ricci curvature of the metric $g$ and $R=g^{\mu\nu}R_{\mu\nu}$. Eq. \eqref{diffusion} is the {\it macroscopic} diffusion equation. The parameter $\sigma>0$ is the diffusion constant. The value $\sigma$ measures the energy transferred from the scalar field to the matter per unit of time due to diffusion.

The fluid correspondence of the diffusion dynamics can be found by projecting the energy-momentum tensor in the direction of $u^{\mu}$ and onto the hypersurface orthogonal to $u^{\mu}$. With this procedure we obtain the energy balance and the Euler equation, respectively,

\begin{equation}\label{cont}
\nabla_{\mu}(\rho u^{\mu})+p\nabla_{\mu}=\sigma n,
\end{equation}
\begin{equation}\label{Euler}
(\rho+p)u^{\mu}\nabla_{\mu} u^{\nu}+u^{\nu}u^{\mu}+\nabla_{\mu}p+g^{\mu\nu}\nabla_{\mu}p=0.
\end{equation}

A viable cosmological model in which dark matter undergoes diffusion in interaction with a dark energy solvent field $\phi$ has been developed in Ref. \cite{VeltenCalogero}, the so called $\phi$CDM model.

For the $\phi$CDM model the energy balances for the matter and the $\phi$ field are described by the system 
\begin{equation}
\dot{\rho}_{dm}+3H(\rho_{dm}+p_{dm})=\sigma n,
\end{equation}
\begin{equation}
\dot{\phi}=-\sigma n.
\end{equation}
In a dimensionless form these equations read
\begin{equation}\label{dimensionlesseq1new}
\frac{d\Omega_m(z)}{dz}=\frac{3\Omega_m(z)}{1+z}-\tilde{\sigma}\frac{(1+z)^2}{E(z)},
\end{equation}
\begin{equation}\label{dimensionlesseq2new}
\frac{d\Omega_\phi(z)}{dz}=\tilde{\sigma}\frac{(1+z)^2}{E(z)},
\end{equation}
where the redshift is written in terms of the scale factor as $z=a^{-1}-1$. We have redefined the diffusion constant to dimensionless units $\tilde{\sigma}=\sigma n_0/3H^3_0$. The dimensionless expasion factor $E(z)=H(z)/H_0$ (subscript $0$ denotes today's values) reads
\begin{equation}\label{dimensionlesseq3new}
E(z)=\sqrt{\Omega_{b}(z)+\Omega_m(z)+\Omega_\phi(z)},
\end{equation} 
The baryoninc component obeys separately to the conservation equation for pressureless fluids which results in $\Omega_b(z) = \Omega_{b0}(1+z)^3$.

For $\tilde{\sigma}=0$ the $\phi$ field remains constant in time and the solution is given by the $\Lambda$CDM model:
\[
\Omega^{(0)}_m(z)=\Omega^{(0)}_{m0}(1+z)^3,\quad\Omega_\phi^{(0)}(z)=\Omega_{\phi0}^{(0)}=1-\Omega^{(0)}_{m0}.
\]

\section{Spherical Collapse Equations}\label{Sec:eq}

Our goal here is to obtain the most general perturbed equations for the evolution of an overdense spherical region collapsing in an expanding universe. We follow Refs. \cite{Rui, Abramo, Abramo2} and references therein. 

Let us first define basic quantities. Note that the spherical collapse happens within the Newtonian framework. For the collapsed region (superscript $c$) one can write
\begin{eqnarray}
\vec{v}^c &=& \vec{u}_0 + \vec{v}_p, \\
\rho^c &=& \rho\left(1+\delta\right) , \\
p^c &=& p + \delta p.
\end{eqnarray}

The velocity of the collapsed region $\vec{v}^c$ can be seen as the balance between the background expansion and the peculiar motion.

The effective expansion rate of the collapsed region is written as
\begin{equation}
h=H+\frac{\theta}{3a},
\end{equation}
where $\theta=\vec{\nabla} \cdot \vec{v}_p$ and $\vec{v}_p$ the peculiar velocity field.

For the collapsing region one has to assure energy conservation. Therefore, each component $i$, possessing an equation of state $p_i=w_i \rho_i$, obeys separately an equation of the type

\begin{equation}\label{eqdeltaorig}
\dot{\delta_i}=-3H(c^2_{eff_i}-w_i)\delta_i-\left[1+w_i+(1+c^2_{eff_i})\delta_i\right]\frac{\theta}{a}
\end{equation}

where the energy density contrast is defined as

\begin{equation}\label{deltadef}
\delta_i = \left(\frac{\delta \rho}{\rho}\right)_i,
\end{equation}
and the effective speed of sound is computed following $c^2_{eff_i} = (\delta p / \delta \rho)_i$.

The dynamics of the perturbed region will be governed by the Raychaudhuri equation
\begin{equation}\label{eqthetaorig}
\dot{\theta}+H\theta+\frac{\theta^2}{3a}=-\frac{a}{2} \sum_i (\delta\rho_i + 3\delta p_i)+a\delta \phi\ .
\end{equation}
Note that the perturbation of the $\phi$ field contributes with a positive sign in the right hand side of this equation in opposition to standard matter fluids.

\subsection{The $\phi$CDM model with $\delta \phi=0$}

Let us in a first moment impose a restriction to our model assuming that $\phi$, which acts as a solvent field, is able neither to cluster nor contribute to form structures, i.e., we treat it as a geometric contribution to the background expansion only. Therefore, we set by hand $\delta \phi=0$ in this analysis. Note that most of the studies in the literature also neglect the importance of the dark energy fluctuations to the clustering process. However, a recent analysis of the impact of dark energy perturbations on the matter clustering can be found in \cite{ronaldo}.

Since the perturbations of the $\phi$ field during the collapse will be neglected our results will depend only on the background expansion. Any difference presented by the $\phi$CDM model when compared to the standard case is due to a non-trivial background expansion caused by the magnitude $\tilde{\sigma}$.

In this case the effective fluids are represented by the baryons and dark matter. Since, $p_b=p_{dm}=0$, both components obey separately an equation of the type (\ref{eqdeltaorig}). Therefore,
\begin{equation} \label{blambda}
\dot{\delta}_b=-\left(1+\delta_b\right)\frac{\theta}{a}, 
\end{equation}
\begin{equation} 
\dot{\delta}_{dm}=-\left(1+\delta_{dm}\right)\frac{\theta}{a}.
\end{equation}

Adapting equation (\ref{eqthetaorig}) to this case, the dynamics of the velocity potential is given by 

\begin{equation} \label{thetalambda}
\dot{\theta}+H\theta+\frac{\theta^2}{3a}=-\frac{a}{2}(\rho_b\delta_b+\rho_{dm}\delta_{dm}).
\end{equation}

For the background we fix $H_0 =72Km/s/Mpc$, $\Omega_{dm0}=0.26$ and $\Omega_{b0}=0.05$. Our final results which will focused on the value of $\tilde{\sigma}$ are not very sensitive to the choice of these background values.

Following \cite{Rui} and \cite{Carames2014} we solve numerically the system of equations (\ref{blambda}) - (\ref{thetalambda}) with initial conditions $\delta_{dm}(z=1000)=3.5\times10^{-3}$, $\delta_b(z=1000)=10^{-5}$ and $\theta (z=1000)=0$. The initial amplitudes of the dark matter and baryonic perturbations exemplify the known fact that baryonic matter tracks the ``already existing'' dark matter potential wells after the decoupling.

In Figs. 1, 2 and 3 we show the results for the dark matter and baryonic growth as well as the expansion rate of the collapsed region $h$, respectively. In the top panels of Figs. 1,2 and 3 the $\Lambda$CDM result is shown with the red solid line. For the dark matter diffusion case we plot dashed lines with values $\tilde{\sigma}=0.01$ and $\tilde{\sigma}=0.05$ which can be identified with the captions of these figures. The bottom panels in Fig. 1 and 2 are complementary to the top ones. They display the relative (percentage) difference in comparison to the $\Lambda$CDM model, i.e., $\Delta \delta_{dm}=(\delta^{\Lambda CDM}_{dm}-\delta^{\tilde{\sigma}}_{dm})/\delta^{\tilde{\sigma}}_{dm}$. From Fig. 3 we can infer the redshift at which the turnaround happens, i.e., $z_{ta}=z(h=0)$. For the $\Lambda$CDM models $z^{\Lambda CDM}_{ta}=0.211$ while for the diffusion model with $\tilde{\sigma}=0.01$ we find $z^{\tilde{\sigma}=0.01}_{ta}=0.154$. If $\tilde{\sigma}=0.01$ the turnaround would occur only in a future time at $z^{\tilde{\sigma}=0.05}_{c}=-0.007$.

We can understand these results uniquely in terms of the background expansion. The interaction between dark matter and the field $\phi$ makes the background expansion faster than in the $\Lambda$CDM case and, consequently, structures form slower in the diffusion scenario.

\begin{figure}\label{fig1}
\begin{center}
\includegraphics[width=0.43\textwidth]{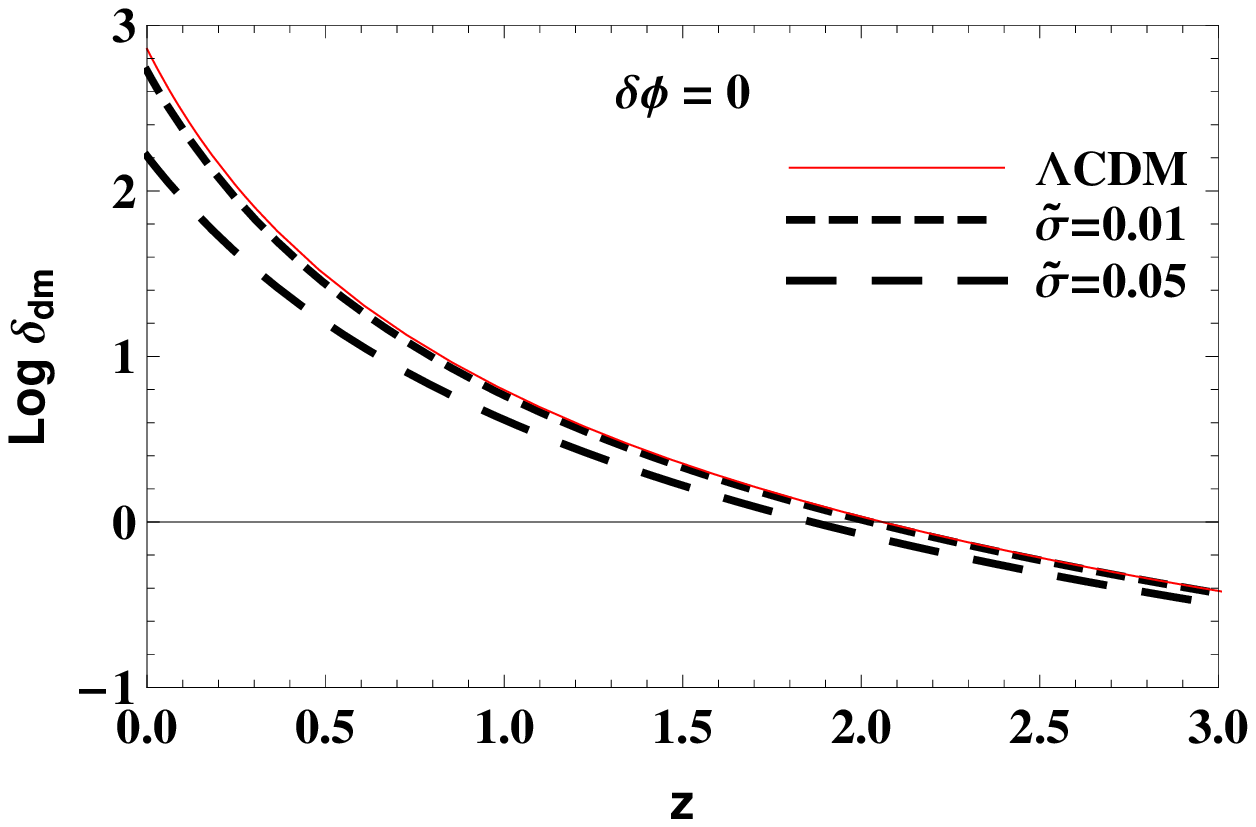}
\includegraphics[width=0.43\textwidth]{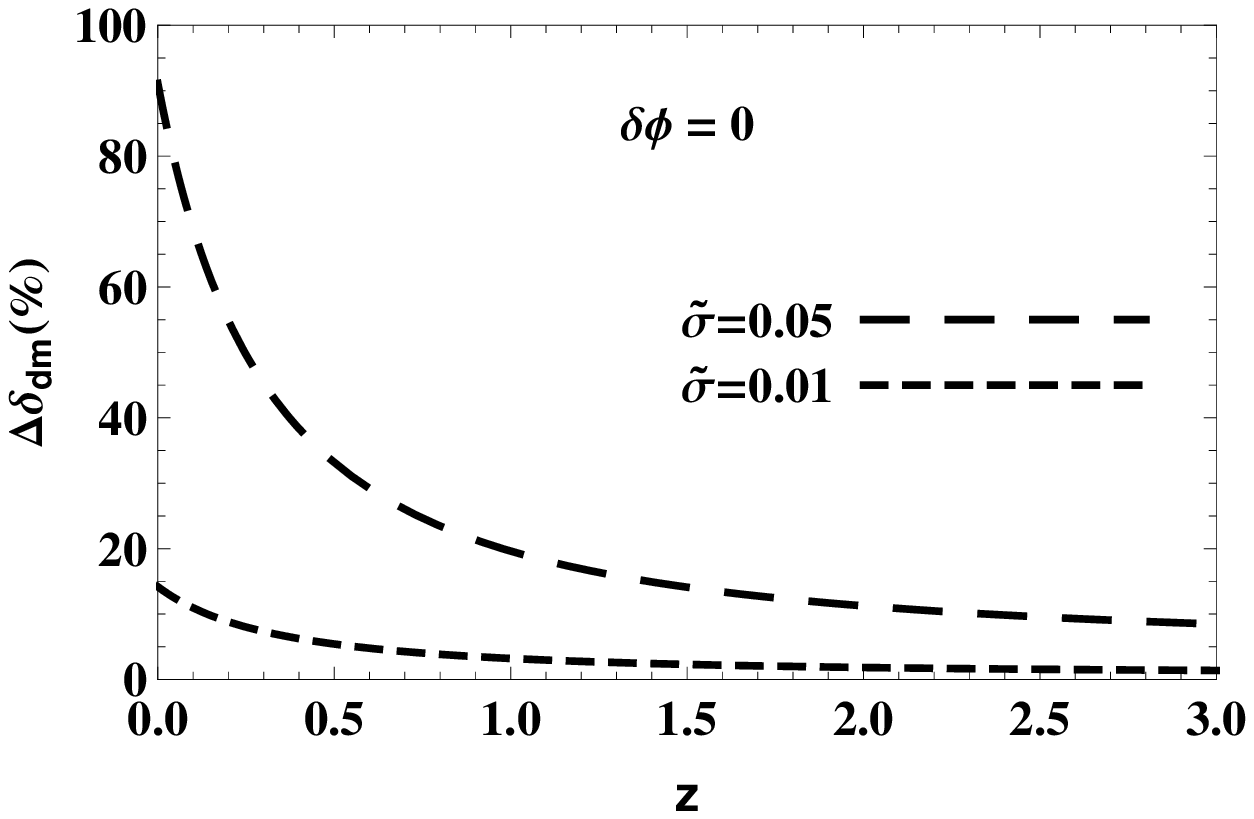}	
\caption{Dark matter perturbation growth as a function of the redshift.}
\end{center}
\end{figure}

\begin{figure}\label{fig2}
\begin{center}

\includegraphics[width=0.43\textwidth]{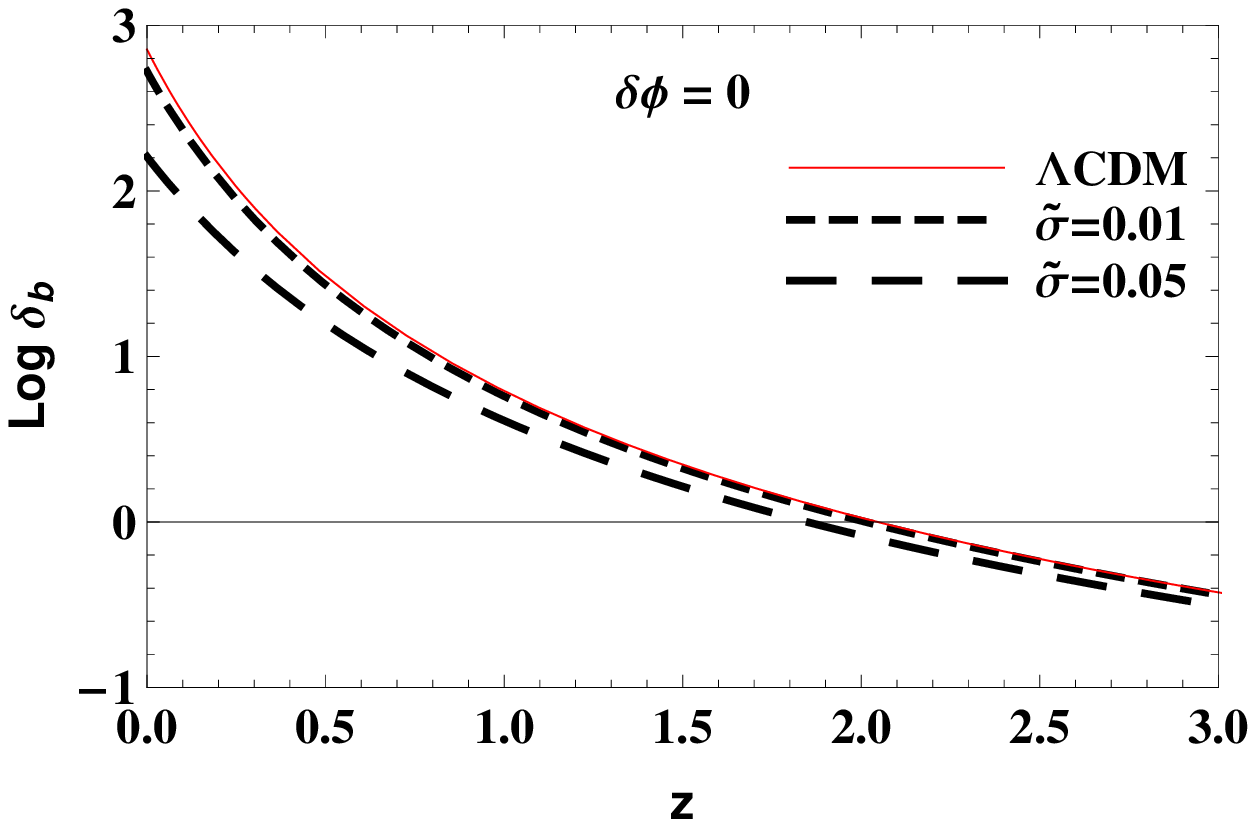}
\includegraphics[width=0.43\textwidth]{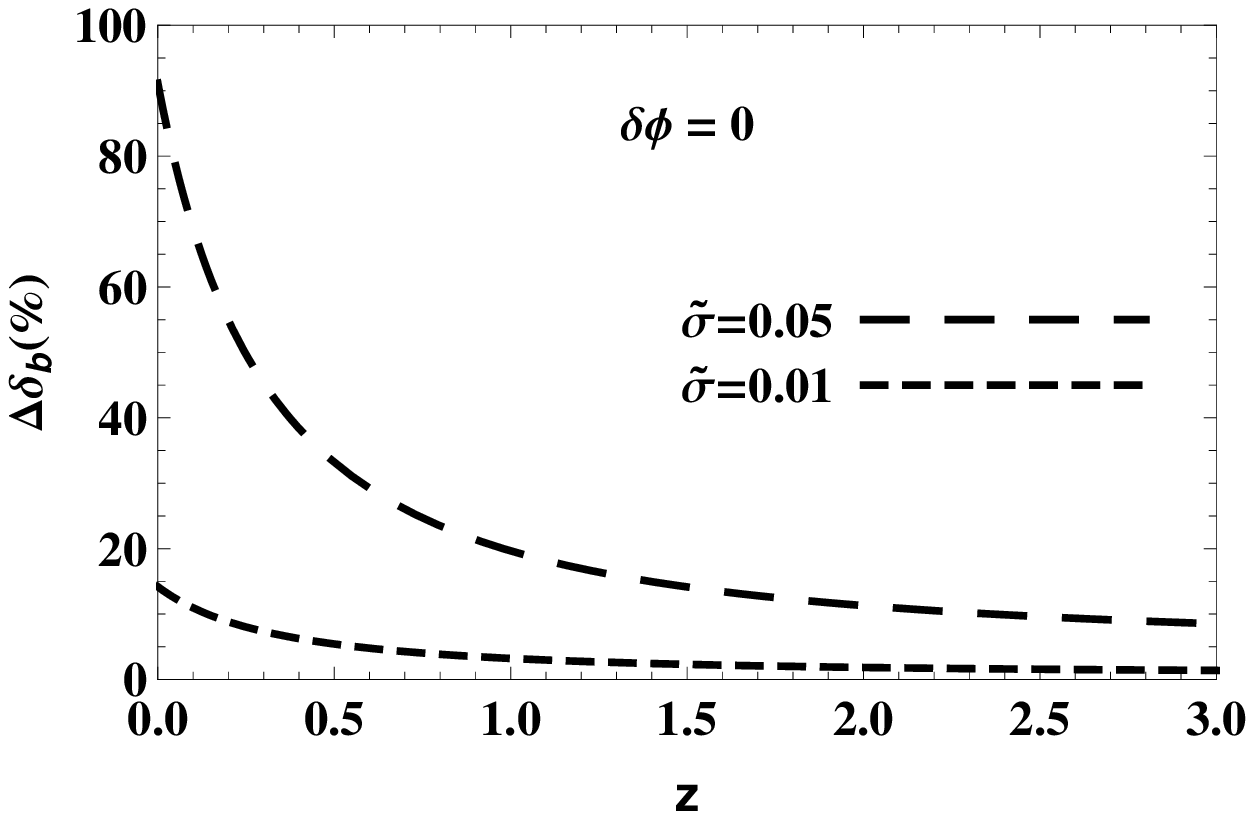}
\caption{Perturbation growth of baryons as a function of the redshift. }
\end{center}
\end{figure}

\begin{figure}\label{fig3}
\begin{center}
\includegraphics[width=0.43\textwidth]{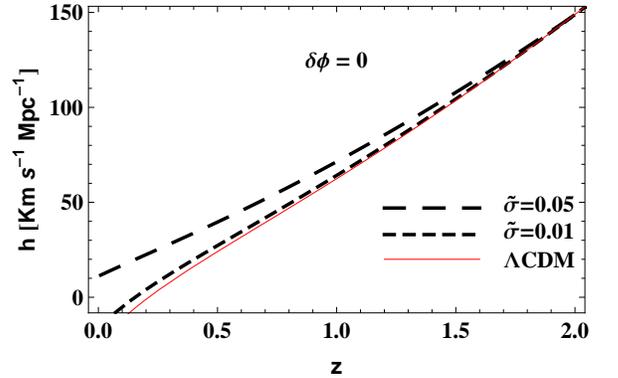}
\caption{Evolution of the expansion rate of the collapsed region.}
\end{center}
\end{figure}

\subsection{Taking into account perturbations of the scalar field: $\delta\phi \neq 0$}

Let us deduce the equations for the spherical collapse when the dark energy perturbation is allowed, i.e., $\delta\phi \neq 0$. The baryonic component is an independent quantity and therefore it obeys the same equation as in the previous case
\begin{equation} \label{b3}
\dot{\delta}_b=-\left(1+\delta_b\right)\frac{\theta}{a}.
\end{equation}

For the dark matter component we have to derive an equivalent equation to (\ref{eqdeltaorig}). Taking into account that the energy balance (\ref{cont}) between the collapsed region (superscript $c$) and the background expansion we find  
\begin{equation}\label{nc}
\dot{\rho}^c_{dm}+3h(\rho^c_{dm}+p^c_{dm})-\dot{\rho}_{dm}-3H(\rho_{dm}+p_{dm})= \sigma n^c-\sigma n.
\end{equation}

Note that $\rho\sim n m$ where $m$ is the particle mass. Therefore, in the same way as we have $\rho^c=\rho(1+\delta_{dm})$, the perturbed particle number can be written as $n^c=n(1+\delta_{dm})$ since $n$ refers to number of dark matter particles. Evaluating (\ref{nc}) we find out that the new terms which are proportional to $\sigma$ cancels out and therefore even in presence of dark energy fluctuations the dark matter obeys to the same equation as in the previous case
\begin{equation}\label{dm3}
\dot{\delta}_{dm}=-\left(1+\delta_{dm}\right)\frac{\theta}{a}. 
\end{equation}

The corresponding version of (\ref{eqthetaorig}) when $\delta\phi \neq0$ becomes
\begin{equation} \label{theta3}
\dot{\theta}+H\theta+\frac{\theta^2}{3a}=-\frac{a}{2}(\rho_b\delta_b+\rho_{dm}\delta_{dm}-2\phi\delta_{\phi})\ .
\end{equation}
Hence we need an equation for the evolution of $\delta_{\phi}$. By deriving with respect to the time the relation $\phi^c=\phi(1+\delta_{\phi})$ we arrive at
\begin{equation}\label{deltaphi}
\phi\dot{\delta}_{\phi} = \dot{\phi}\left(\delta_{dm}-\delta_{\phi}\right).
\end{equation}

The system (\ref{b3}), (\ref{dm3}), (\ref{theta3}) and (\ref{deltaphi}) will be solved numerically with the same initial conditions as in the previous analysis but adding $\delta_{\phi}(z=1000)=3.5\times10^{-3}$. The results are shown in Figs. 4, 5, 6 and 7. Fig. 4 can be directly compared to Fig. 1 (top panel), Fig.6 corresponds to the same as Fig. 2 (top panel) and Fig. 7 to the 3, respectively. 

As seen in Fig. 4, allowing for the perturbation of the dark energy field the dark matter growth is clearly favoured. Here, in oppostition to the case $\delta_{\phi}=0$, the background expansion plays a secondary role. The dynamics of $\delta_{dm}$ is mostly driven by the energy flux comming from the scalar field perturbation $\delta_{\phi}$ which, on the other hand, evolves to form underdense regions.  

Since perturbations of the scalar field are also able to modify the total first order dynamics of the collapsing region we define the quantity

\begin{equation}
\Delta_{T}=\frac{\Omega_{dm} \delta_{dm}+\Omega_{\phi}\delta_{\phi}}{\Omega_{dm}}.
\end{equation}

and plot it in Fig.5. This quantity tries to illustrate the effective contribution of the scalar field perturbation to the dark matter one. In some sense, it can be seen as a possible representation of the total ``dark'' sector perturbation within the shell. Since $\Delta_{T}$ is suppressed in comparison to $\delta_{m}$ we interpret it as a $\delta_{\phi}$ forming an underdense region, i.e., a void. In particular, as long the perturbation remain below the non-linear regime $\Delta_{T}<1$ we observe a similar growth supression effect as in the case $\delta_{\phi}=0$. However, the total dark perturbation growth represented by $\Delta_{T}$ is propelled when the nonlinear regime is crossed. This new effect was not observed in the previous analysis and even in the linear theory studied in \cite{VeltenCalogero}.

The baryonic growth shown in Fig. 6 tracks the dark matter one rather than $\Delta_{T}$. Also, the faster collapse of the expansion $h$ shown in Fig. 7 is consistent with the amplitude excess of $\delta_{dm}$ and $\delta_b$.

\begin{figure}\label{fig4}
\begin{center}
\includegraphics[width=0.43\textwidth]{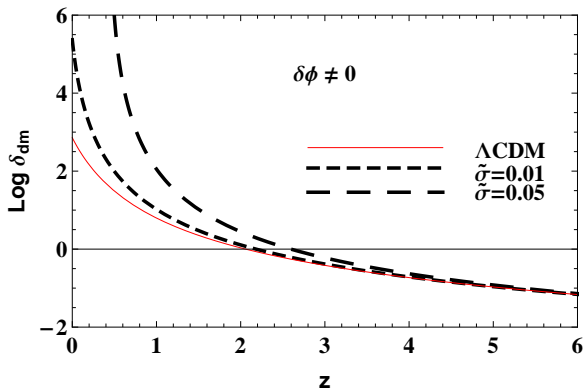}
\caption{Dark matter perturbation growth as a function of the redshift.}
\end{center}
\end{figure}

\begin{figure}\label{fig4a}
\begin{center}
\includegraphics[width=0.43\textwidth]{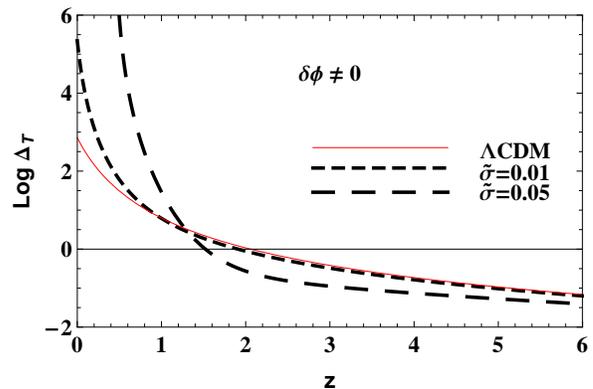}
\caption{Perturbation growth of the quantity $\Delta_{T}$ as a function of the redshift.}
\end{center}
\end{figure}

\begin{figure}\label{fig5}
\begin{center}
\includegraphics[width=0.43\textwidth]{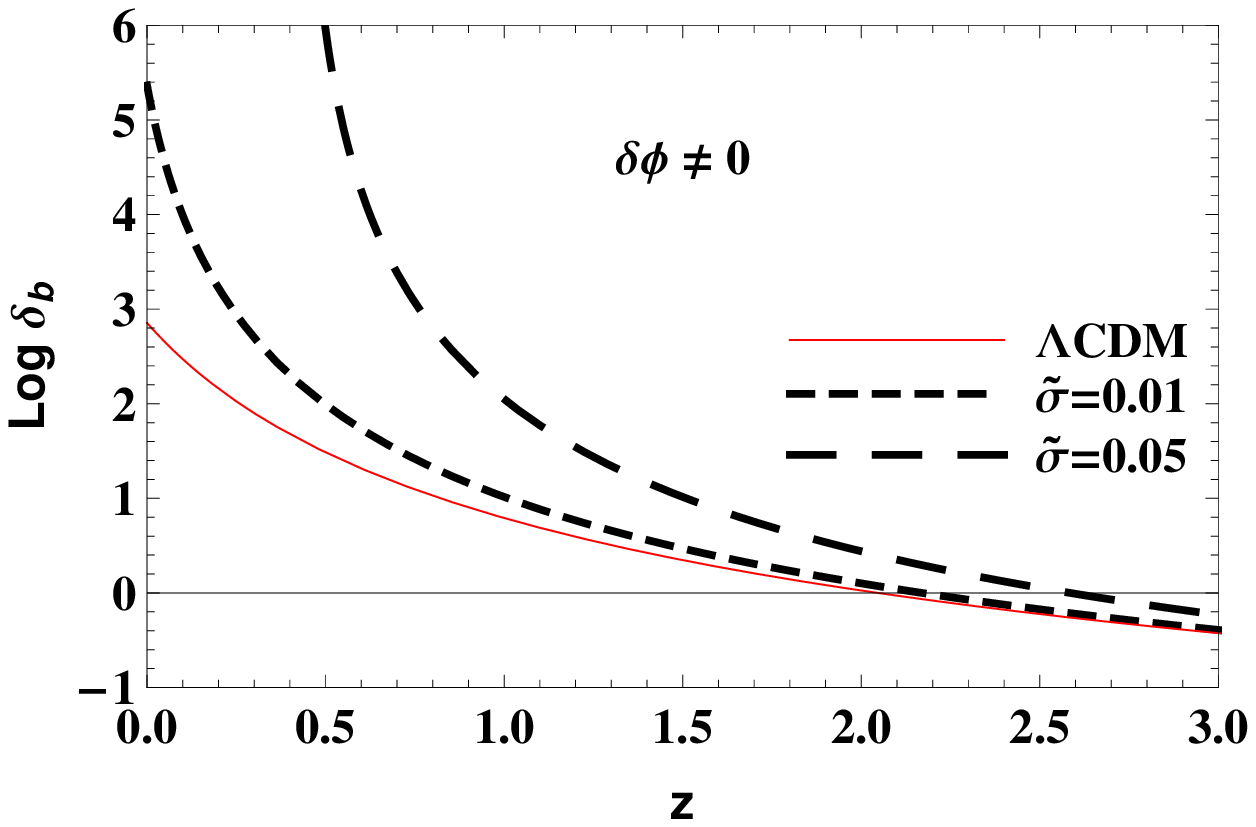}
\caption{Perturbation growth of baryons as a function of the redshift.}
\end{center}
\end{figure}

\begin{figure}\label{fig6}
\begin{center}
\includegraphics[width=0.43\textwidth]{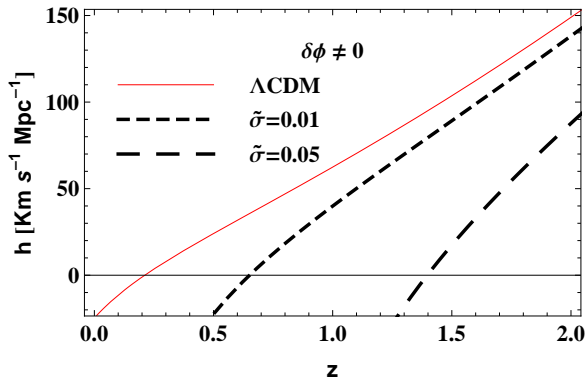}
\caption{Evolution of the expansion rate of the collapsed region.}
\end{center}
\end{figure}

\section{Conclusions}

We have studied the spherical top-hat collapse when dark matter particles undergo velocity diffusion. This approach is justified here since no pressure gradients are present in the model. The dark matter diffusion process is compatible with general relativity since a cosmological solvent field is added to the field equations. The resulting FLRW expansion corresponds to an interacting model where energy flows from the dark energy solvent field $\phi$ to the dark matter component at a rate which is proportional to the coefficient $\sigma$. The $\Lambda$CDM model is recovered when $\sigma=0$.

The linear perturbation theory, in particular the analysis of the matter power spectrum, indicates that the agreement with the large scale structure do exist for values $\tilde{\sigma}<0.01$ \cite{VeltenCalogero}.

Within the dark matter diffusion model we have studied two subcases, namely, when $\delta \phi = 0$ and $\delta \phi \neq 0$. The reason for choosing the former case consists in admmiting the $\phi$ field as a geometric quantity which, of course, does not cluster and therefore is only able to modify the collapse via its effects on the background expansion. Indeed, this is a case similar to the cosmological constant. In general, it is expected that perturbations in the dark energy sector are irrelevant in comparison to the matter ones. For this case the value $\tilde{\sigma}=0.01$ shows only a slightly growth suppression when compared to the standard cosmology. 

The second subcase studied ($\delta \phi \neq 0$) represents a more complete description of the perturbative evolution of the model. In general, the dynamics is much more sensitive for the same values of $\tilde{\sigma}$. Now the total perturbation within the collapsing region can be splitted into the two ``dark''components. In order to illustrate the behavior of the scalar field perturbation we have defined the quantity $\Delta_{T}$ from which one can infer that $\delta_{\phi}$ form underdense regions.

Concerning the magnitude of the diffusion mechanism we conclude that the study of the dark matter top-hat collapse under diffusion provides us complementary results concerning the evolution of the matter fluctuations in an expanding universe. In some sense, our results corroborate previous findings about the magnitude of the dark matter diffusion. In general, this qualitative analysis allows us to conclude that diffusive dark matter collapses in a very similar way as the standard model for $\tilde{\sigma}<0.01$. We hope that in a future contribution a more detailed investigation using for example the mass functions of the model can be used to further explore the parameter space assessing the region $\tilde{\sigma} \lesssim 0.001$.

\textbf{Acknowledgement}:  We thank CNPq (Brazil) for partial financial support.

\end{document}